# On "Strong control, conservative point estimation and simultaneous conservative consistency of false discovery rates":

## Does a large number of tests obviate confidence intervals of the FDR?

### David R. Bickel*

*Office of Biostatistics and Bioinformatics*
*Medical College of Georgia*
*Augusta, GA 30912-4900*

*bickel@prueba.info*
*www.davidbickel.com*

March 27, 2004

**Summary.**
A previously proved theorem gives sufficient conditions for an estimator of the false discovery rate (FDR) to conservatively converge to the FDR with probability 1 as the number of hypothesis tests increases, even for small sample sizes. It does not follow that several thousand tests ensure that the estimator has moderate variance under those conditions. In fact, they can hold even if the test statistics have long-range correlations, which yield unacceptably wide confidence intervals, as observed in genomic data when there are 8 or 16 individuals (microarrays) per group. Thus, informative FDR estimation will include some measure of its reliability.

*Address after March 31, 2004: Pioneer, Bioinformatics and Discovery Research, 7250 NW 62nd Ave., P. O. Box 552, Johnston, IA 50131-0552

# 1 Introduction

In fields as diverse as astronomy, the mining of commercial data, and genomics, researchers may test dozens, hundreds, or thousands of hypotheses at a time, leading to the problem of multiple comparisons. Addressing this in the *Journal of the Royal Statistical Society*, Benjamini and Hochberg (1995) defined the false discovery rate (FDR) to equal the expected proportion of rejected null hypotheses that are true if the probability of rejection is positive or to equal 0 if null hypotheses are almost never rejected. Benjamini and Hochberg (1995) also provided a way to control the FDR under the independence of test statistics. In a more recent issue, Storey, Taylor, and Siegmund (2004) provided informative theorems on the estimation as well as the control of the FDR. Storey, Taylor, and Siegmund (2004) proved that if the p-values, and thus the test statistics, satisfy certain conditions, including "weak dependence," then, as $m$, the number of null hypotheses, goes to infinity, a conservative estimator of the FDR is

$$\hat{\nabla}(\alpha) = \hat{\pi}_0 \, \hat{F}_0(\alpha) / \hat{F}(\alpha), \tag{1}$$

where $\hat{\pi}_0$ is a conservative estimator of $\pi_0$, the proportion of null hypotheses that are true, $\hat{F}_0(\alpha)$ is the empirical distribution of test statistics under the null hypothesis, $\hat{F}(\alpha)$ is the marginal empirical distribution of test statistics, and $\alpha$ is the significance level, the test-wise Type I error rate. (A null hypothesis is rejected if and only if its p-value is less than or equal to $\alpha$.) If the null distribution, $F_0(\alpha)$, is known, it may be used in place of $\hat{F}_0(\alpha)$. Storey, Taylor, and Siegmund (2004) also pointed out that the FDR control method of Benjamini and Hochberg (1995) is asymptotically equivalent to rejecting as many null hypotheses as possible subject to holding $\hat{\nabla}(\Gamma)$ under some value; Benjamini and Hochberg (1995) effectively used $\hat{\pi}_0 = 1$. Letting $R(\alpha)$ be the number of rejections, the estimator corresponding to this conservative choice and to uniform $F_0(\alpha)$ is

$$\hat{\nabla}_1(\alpha) = \frac{\alpha}{R(\alpha)/m}, \tag{2}$$

which will be used to confine stochasticity to the denominator, without loss of generality.

Given the alternative hypothesis distribution and empirical distribution, $F_1(\alpha)$ and $\hat{F}_1(\alpha)$, Storey, Taylor, and Siegmund (2004) defined "weak dependence" by three conditions, the most restrictive of which is

$$\begin{aligned} \forall_{\alpha \in (0,1]} \Pr\left(\lim_{m \to \infty} \hat{F}_0(\alpha) = F_0(\alpha)\right) = 1; \\ \forall_{\alpha \in (0,1]} \Pr\left(\lim_{m \to \infty} \hat{F}_1(\alpha) = F_1(\alpha)\right) = 1. \end{aligned} \tag{3}$$

Since not all types of dependence satisfying equations (3) and the other two conditions would typically be considered weak, this type of dependence will be called *Storey dependence*, after Storey (2002). An example of "strong" Storey dependence is provided in the next section, dependence that questions how large $m$ must be to approximate the limits of equations (3). While equations (3) do not depend on the sample size, it will be seen that $\hat{\nabla}(\alpha)$ can lack even minimal reliability as an estimator of the FDR for small sample sizes, in spite of $m \approx 10{,}000$ and the important findings of Storey, Taylor, and Siegmund (2004). Thus, even a very large value of $m$ cannot substitute for a sufficiently large sample size.

# 2 Reliability of FDR estimation under Storey dependence

## 2.1 Long-range correlations

The dependence structure of the test statistics may be studied using a simple model of the variance of $\hat{F}(\alpha)$:

$$\begin{aligned} \operatorname{var} m \hat{F}(\alpha) &= \operatorname{var} R = F(\alpha)(1 - F(\alpha)) m^{2H(\alpha)}; \\ \operatorname{var} \hat{F}(\alpha) &= \frac{F(\alpha)(1 - F(\alpha))}{m^{2(1 - H(\alpha))}}. \end{aligned} \tag{4}$$

Geophysicists and statistical physicists call $H(\alpha)$ the Hurst exponent or Hurst parameter; $0 < H(\alpha) < 1$. Equation (4) never holds exactly for physical phenomena, but describes the approximate scaling in the variance for a large class of complex processes, including those for which $m$ is a discrete time scale or spatial resolution. In the case of independence ($\forall_{\alpha \in (0,1]} H(\alpha) = 1/2$), var $m \hat{F}(\alpha)$ is the variance of a binomial process with $m$ trials and probability $F(\alpha)$ of success. $H(\alpha) < 1/2$ corresponds to negative dependence, whereas $H(\alpha) > 1/2$ corresponds to positive dependence. A process satisfying equation (4) and $H(\alpha) > 1/2$ is said to have *long-range correlations* in the sense that its cumulative autocorrelation function does not converge as more terms are added. This type of dependence is strong: the autocorrelation decays much slower than exponentially. Nonetheless, since var $\hat{F}(\alpha) \to 0$ as $m \to \infty$ and since empirical distributions are asymptotically unbiased, equations (3) are satisfied.

The effect of $H(\alpha)$ on the reliability of FDR estimation may be quantified by confidence intervals. Inasmuch as $\hat{F}(\alpha)$ is unbiased and approximately normal, a 95% confidence interval of the FDR estimator $\hat{\nabla}_1(\alpha)$ is

$$\begin{aligned} \text{CI}_1(95\%; \alpha) &= \left( \frac{\alpha}{\Phi^{-1}(97.5\%; F(\alpha), \text{var } \hat{F}(\alpha))}, \frac{\alpha}{\Phi^{-1}(2.5\%; F(\alpha), \text{var } \hat{F}(\alpha))} \right) \\ &\approx \left( \frac{\alpha}{F(\alpha) + (1.96)\sqrt{\text{var } \hat{F}(\alpha)}}, \frac{\alpha}{F(\alpha) - (1.96)\sqrt{\text{var } \hat{F}(\alpha)}} \right); \end{aligned} \tag{5}$$

$\Phi^{-1}(\bullet; \mu, \sigma^2)$ is the quantile function for $N(\mu, \sigma^2)$. As $\hat{\nabla}_1(\alpha)$ is a conservative estimator of the FDR, approximately 95% of the confidence intervals capture $F_0(\alpha)/F(\alpha)$, rather than the FDR, $\pi_0 F_0(\alpha)/F(\alpha)$. In that sense, they may be considered *conservative confidence intervals of the FDR*. Such intervals for $\alpha = 0.01$, $H(0.01) \geq 0.5$, $m = 10{,}000$, and $F(0.01) = 0.2$ are plotted in Fig. 1, which demonstrates that the FDR cannot be reliably estimated if $H(\alpha)$ is too high, even though Storey dependence holds.

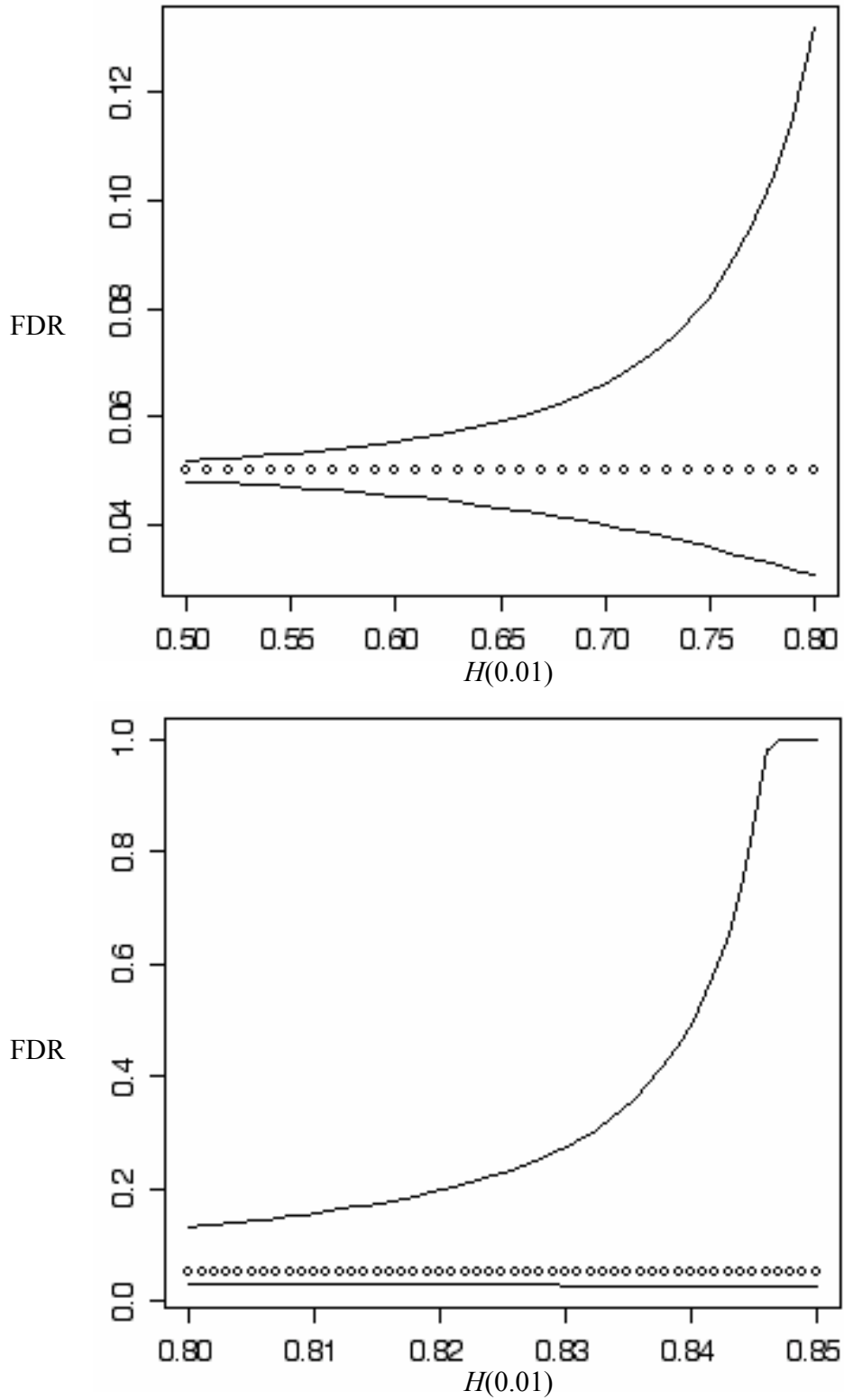

**Fig. 1.** Conservative FDR estimator and its confidence intervals for different degrees of dependence at the 0.01 significance level, with $H(0.01)=0.50$ in the case of independence.

## 2.2 Bootstrap inference

For a particular data set, the reliability of FDR estimation may be judged using $\hat{\sigma}_B(\alpha)$, the sample standard deviation of $R(\alpha)/m$ over $B$ bootstrap samples, as an estimate of $\sqrt{\operatorname{var}\hat{F}(\alpha)}$. The corresponding bootstrap approximation to the conservative 95% confidence interval of the FDR (5) is

$$\mathrm{CI}_1(95\%;\ \alpha,\ B) = \left(\frac{\alpha}{R(\alpha)/m + (1.96)\,\hat{\sigma}_B(\alpha)},\ \frac{\alpha}{R(\alpha)/m - (1.96)\,\hat{\sigma}_B(\alpha)}\right). \tag{6}$$

This calculation holds even without assuming long-range correlations, but when equation (4) does roughly approximate the dependence structure, $H(\alpha)$ is naturally estimated by

$$\hat{H}(\alpha;B) = 1 - \log\left(\frac{\sqrt{R(\alpha)\,(1 - R(\alpha)/m)/m}}{\hat{\sigma}_B(\alpha)}\right)\bigg/\log m. \tag{7}$$

# 3 Dependence in gene expression

The theoretical study of the last section is complemented by a study of gene expression microarrays of Yeoh *et al.* (2002). The data set includes oligonucleotide microarrays from children with various subtypes of leukemia, the largest two of which are the hyperdiploid > 50 chromosomes subgroup (H50, 64 children) and the TEL-AML1 subgroup (79 children). The microarray of each child has an *expression value*, a measured level of mRNA abundance, for each of 12,625 genes, for a total of $12{,}625 \times (64 + 79)$ expression values in the two subgroups. A p-value of differential expression between them is computed for each gene using the two-sided Wilcoxon rank-sum test after normalizing each microarray by its median expression value, so $m = 12{,}625$ and the sample sizes are 64 and 79. Two separate resampling procedures are described below: 1. bootstrapping to estimate a confidence interval of the FDR for the entire samples and 2. resampling without replacement to assess the effect of smaller sample sizes on confidence intervals of the FDR.

## 3.1 Confidence intervals for full samples

Each of 100 bootstrap samples was generated by randomly selecting, with replacement, 64 microarrays from the H50 subgroup and 79 microarrays from the TEL-AML1 subgroup, yielding $12{,}625 \times 100$ bootstrap p-values in addition to the 12,625 p-values of the original data. Each null hypothesis was rejected if its p-value was not greater than $\alpha$, with the results of Table 1 from equations (2), (6) and (7). The tight confidence intervals indicate reliable FDR estimation notwithstanding the strong violation of independence $\left(\hat{H}(\alpha; 100) \gg 0.500\right)$. Unfortunately, the computations below show that same cannot be said for much smaller sample sizes.

| $\alpha$ | $\hat{V}_1(\alpha)$ | $CI_1(95\%; \alpha, 100)$ | $\hat{H}(\alpha; 100)$ |
|---|---|---|---|
| **0.01** | 0.052 | (0.045, 0.062) | 0.658 |
| **0.05** | 0.177 | (0.158, 0.201) | 0.656 |

**Table 1.** Conservative FDR estimates, conservative 95% confidence intervals, and dependence exponent estimates for gene expression data of sample sizes 64 and 79.

The confidence intervals are valid whether or not equation (4) is appropriate. However, in light of the finding that eukaryotic regulatory networks can involve thousands of genes (Lee *et al.* 2002), it is likely that their test statistics have long-range correlations, and thus that equation (4) with $H > 1/2$ accurately describes their dependence structure.

## 3.2 Smaller sample sizes

To examine the reliability of FDR estimation for sample sizes of $n$ microarrays per group, each of 100 subsamples was generated by randomly selecting, without replacement, $n$ microarrays from the H50 subgroup and $n$ microarrays from the TEL-AML1 subgroup, yielding $12{,}625 \times 100$ subsample p-values for each value of $n$. As with the full sample, equations (2), (6) and (7) were used to compute the estimates and confidence intervals, except with $\hat{\mu}_{n,100}(\alpha)$ and $\hat{\sigma}_{n,100}(\alpha)$, the sample mean and sample standard deviation of $R/m$ across 100 subsamples, in place of $R/m$ and $\hat{\sigma}_{100}(\alpha)$, respectively:

$$CI_1(95\%; \alpha, n, 100) = \left( \frac{\alpha}{\hat{\mu}_{n,100}(\alpha) + (1.96)\,\hat{\sigma}_{n,100}(\alpha)},\ \frac{\alpha}{\hat{\mu}_{n,100}(\alpha) - (1.96)\,\hat{\sigma}_{n,100}(\alpha)} \right); \qquad (8)$$

$$\hat{H}(\alpha; n, 100) = 1 - \log\left( \frac{\sqrt{\hat{\mu}_{n,100}(\alpha)\,(1 - \hat{\mu}_{n,100}(\alpha))}}{\hat{\sigma}_{n,100}(\alpha)} \right) \Big/ \log m. \qquad (9)$$

Table 2 quantifies the lack of reliability in FDR estimation for sample sizes typical of current microarray experiments.

| $n$ | $\alpha$ | $\hat{\mu}_{n,100}(\alpha)$ | $CI_1(95\%; \alpha, n, 100)$ | $\hat{H}(\alpha; n, 100)$ |
|---|---|---|---|---|
| 8 | 0.01 | 0.292 | (0.169, 1.000) | 0.718 |
| 8 | 0.05 | 0.451 | (0.320, 0.765) | 0.724 |
| 16 | 0.01 | 0.129 | (0.093, 0.212) | 0.700 |
| 16 | 0.05 | 0.333 | (0.262, 0.454) | 0.700 |

**Table 2.** Conservative mean FDR estimates, conservative 95% confidence intervals, and dependence exponent estimates for gene expression data.

## 4 Conclusions

The data analysis indicates that FDR estimation in gene expression studies can be very misleading unless confidence intervals are reported with estimates, especially when there are less than 16 independent microarrays per group. This, with the fact that long-range dependence is a type of Storey dependence, emphasizes the need to distinguish the theorems of Storey, Taylor, and Siegmund (2004) from the naive conclusion that FDR estimation will not suffer much from small sample sizes as long as the number of tests is large.

Even when the FDR is controlled rather than estimated, the reliability of estimation can impact the interpretation of the results of FDR control, given the close connection between FDR estimation and FDR control (Storey, Taylor, and Siegmund 2004). For example, reporting an FDR estimate with a large confidence interval could help prevent a non-statistician from believing that about 5% of discoveries are false on the basis of the fact that the FDR was controlled at the 5% level.

Bootstrapping to obtain confidence intervals is not always practical, as when the null distribution is obtained by random permutations with limited computer resources. In such cases, rough 95% conservative confidence intervals of the FDR may be obtained from $H = 0.7$ with equations (4) and (5):

$$\left( \frac{\alpha}{R(\alpha)/m + (1.96)\, m^{-0.8} \sqrt{R(\alpha)\,(1 - R(\alpha)/m)}}, \right.$$
$$\left. \frac{\alpha}{R(\alpha)/m - (1.96)\, m^{-0.8} \sqrt{R(\alpha)\,(1 - R(\alpha)/m)}} \right). \tag{10}$$

Although such intervals are not as accurate as those obtained from the data at hand (6), they are preferable to omitting any indication of the reliability of FDR estimates.